\documentclass[twocolumns]{aa}

\usepackage{graphicx}
\usepackage{multirow}
\usepackage{times}
\begin{document}

   \title{Sensitivity of solar off-limb line profiles to electron density
   stratification and the  velocity distribution anisotropy}

   \author{N.-E. Raouafi\inst{1}\fnmsep\inst{2}
          \and
          S. K. Solanki\inst{1}}
   \offprints{N.-E. Raouafi}

   \institute{Max-Planck-Institut f\"ur Sonnensystemforschung,
              Max-Planck-Stra$\ss$e 2, 37191 Katlenburg-Lindau, Germany
	   \and
	 National Solar Observatory, 950 North Cherry Avenue, Tucson, AZ 85726, USA \\
         \email{nraouafi@nso.edu ; solanki@mps.mpg.de}}

\authorrunning{Raouafi \& Solanki}
   \date{Received ; accepted }

   \abstract{The effect of the electron density stratification on the intensity
profiles of the \ion{H}{i} Ly-$\alpha$ line and the \ion{O}{vi} and \ion{Mg}{x}
doublets formed in solar coronal holes is investigated. We employ an analytical
2-D model of the large scale coronal magnetic field that provides a good
representation of the corona at the minimum of solar activity. We use the 
mass-flux conservation equation to determine the outflow speed of the solar wind
at any location in the solar corona and take into account the integration along
the line of sight (LOS). The main assumption we make is that no anisotropy in
the kinetic temperature of the coronal species is considered. We find that at
distances greater than 1~$R_{\sun}$ from the solar surface the widths of the
emitted lines of \ion{O}{vi} and \ion{Mg}{x} are sensitive to the details of the
adopted electron density stratification. However, Ly-$\alpha$, which is a pure
radiative line, is hardly affected. The calculated total intensities of
Ly-$\alpha$ and the \ion{O}{vi} doublet depend to a lesser degree on the density
stratification and are comparable to the observed ones for most of the
considered density models. The widths of the observed profiles of Ly-$\alpha$
and \ion{Mg}{x} are well reproduced by most of the considered electron density
stratifications, while for the \ion{O}{vi} doublet only few stratifications give
satisfying results. The densities deduced from SOHO data result in \ion{O}{vi}
profiles whose widths and intensity ratio are relatively close to the values
observed by UVCS although only isotropic velocity distributions are employed.
These density profiles also reproduce the other considered observables with good
accuracy. Thus the need for a strong anisotropy of the velocity
distribution (i.e. a temperature anisotropy) is not so clear cut as previous
investigations of UVCS data suggested. However, these results do not rule
completely out the existence of some degree of anisotropy in the corona. The
results of the present computations also suggest that the data can also be
reproduced if protons, heavy ions and electrons have a common temperature, if
the hydrogen and heavy-ion spectral lines are also non-thermally broadened by a
roughly equal amount.

 }

   \maketitle

\markboth{Raouafi \& Solanki: Density Stratification Effect on Solar Off-limb Line
Profiles}{Raouafi \& Solanki: Density Stratification Effect on Solar Off-limb Line
Profiles}

\section{Introduction}

Extreme-ultraviolet (EUV) spectral lines provide valuable information on the
plasma conditions in the solar wind acceleration zone. Information on the
densities, the temperatures, the bulk velocities and the velocity distributions
of different species, can be obtained from the total intensity, Doppler width
and shift of multiple line profiles with the appropriate instrument. Basically
three techniques are used to deduce both bulk velocity and the velocity
distribution.

The first is the usual Doppler effect that provides information on the
LOS velocity of the emitting source. The second is the Doppler dimming
which yields information on (microscopic and macroscopic) velocity components
not directed along the LOS, on the basis of the technique proposed by Rompolt
(1967 \& 1969) and Hyder \& Lites (1970). See, e.g., Beckers \& Chipman
(1974) for a description. Note that Doppler dimming acts only on the radiative
component of a line. Rompolt (1967, 1969 \& 1980), Heinzel \& Rompolt (1987)
and Gontikakis et al. (1997a,b) used the Doppler dimming of Ly-$\alpha$ to study
moving prominences. It has later been considered to diagnose the solar wind
(Kohl \& Withbroe 1982; Withbroe et al. 1982; Strachan et al. 1989 \& 1993;
etc.).

Optical pumping is a third process. It happens when, due to the Doppler
effect, the absorption profile of the moving atom overlaps with a neighboring
incident spectral line. This is the case for the coronal \ion{O}{vi}
1037.61~{\AA} line that can be excited by the chromospheric \ion{C}{ii}
doublet at 1036.3367 and 1037.0182~{\AA} when the solar wind speed reaches
values between $\sim100$ and $\sim500$ km~s$^{-1}$ (Noci et al. 1987; Kohl
et al. 1998; Li et al. 1998). Finally,  Cranmer et al. (1999a) used optical
pumping by \ion{Fe}{iii} 1035.77~{\AA} to diagnose wind speeds around $\sim530$
km~s$^{-1}$.

UVCS (the UltraViolet Coronagraph Spectrometer; Kohl et al. 1995 \& 1997) on the
SOHO (the SOlar and Heliospheric Observatory; Domingo et al. 1995) mission
yielded valuable data on the solar corona, including the polar coronal holes up
to more than 3.5~$R_{\sun}$ (Kohl et al. 1997 \& 1998; Noci et al. 1997; Habbal
et al. 1997; Li et al. 1998; etc.). One of the most exciting results obtained
from data recorded by UVCS concerns the very broad lines emitted by heavy ions
(namely \ion{O}{vi} and \ion{Mg}{x}) in polar coronal holes during the minimum
of the solar activity cycle. From the analysis of profiles of coronal lines
(mainly the Ly-$\alpha$, \ion{O}{vi} and \ion{Mg}{x} doublets) Kohl et al. (1997
\& 1998), Noci et al. (1997) and others concluded that the evidence for highly
anisotropic velocity distributions of the cited species is strong. For
\ion{O}{vi} and \ion{Mg}{x} the ratio of kinetic temperatures in the directions
perpendicular and parallel to the coronal magnetic field, respectively, is found
to range from 10 to more than 100 (Kohl et al. 1997 \& 1998; Cranmer et al.
1999b; etc.). In addition, the heavy ions are deduced to be significantly hotter
and faster than the protons (Li et al. 1997).

In the present paper we extend the work of Raouafi \& Solanki (2004; hereafter
abbreviated as RS04) in different ways. Firstly, we consider a more complete set
of published density profiles. Secondly, we study the effect of the density
stratification on spectral lines having very different formation mechanisms:
Ly-$\alpha$ is purely radiative; \ion{Mg}{x} lines are exclusively collisional
and the \ion{O}{vi} doublet results from both mechanisms. In addition, we also
calculate the LOS integrated total intensities of the Ly-$\alpha$ and
\ion{O}{vi} lines.

\section{Line formation in the corona}

In an optically thin medium, the general form of the contribution to the
emissivity of a given point on the LOS, assuming a Maxwellian velocity
distribution and a Gaussian shaped incident profile, is given by
\begin{eqnarray}
\begin{array}{l}
{\displaystyle 
{\cal{I}}(\nu)= \int_{LOS} {\rm{d}}Z \;\Bigg\{
\frac{N_{X^{n+}}(\theta)\,{N_{\rm e}}(\theta)\,{\alpha}_{lu}}
{4\,\pi\,\sqrt{\pi}\,\alpha_s} \,
{\rm{e}}^{-\left(\frac{\delta\nu-\frac{\nu_0}{c}u_Z}{\alpha_s}\right)^2}  }\\
\;\;\;+ {\displaystyle 
\frac{N_{X^{n+}}(\theta)\,B_{lu}\,{\cal{I}}_{R}}{\left(4\,\pi\right)^2} } 
 {\displaystyle
\int_{\Omega}\!{\rm{d}}\Omega\;\left[f \times
{\cal{D}} \times {\cal{P}} \times {\cal{M}}\right](\Omega,\theta) \Bigg\}};
\label{Gen_Emis_Formula}
\end{array}
\end{eqnarray}
where $N_{X^{n+}}(\theta)$ and ${N_{\rm{e}}}(\theta)$ are densities of
scattering ions $X^{n+}$ and electrons (cm$^{-3}$), respectively. $B_{lu}$ and
${\alpha}_{lu}$ are the Einstein coefficient for absorption between the two
atomic levels $l$ and $u$\footnote{In the present paper, we consider the
assumption of a two level atom.} and the coefficient of electronic collisions,
respectively. The width of the microscopic velocity distribution $\alpha_s$ is
assumed to be isotropic ({\it{no anisotropy in the kinetic temperature of
scattering atoms/ions is considered throughout this paper)}}). $\alpha_s$
includes the contribution due to thermal motions.
${\displaystyle\delta\nu=\nu-\nu_0}$, where $\nu_0$ is the rest frequency of the
reemitted line. The geometry of the scattering process is given by Fig.~1 in
RS04. The other terms entering Eq.~(\ref{Gen_Emis_Formula}) are defined in the
same paper. The different values of the coefficient ${\cal{W}}_{ul}$ in the
expression of ${\cal{M}}(\Omega,\theta)$ and the corresponding spectral lines
studied here are given in Table~\ref{wul_values}. Eq.~(\ref{Gen_Emis_Formula})
is general and applies to any spectral line irrespective of whether it is formed
by collisions and/or by radiation. An exception is when the incident profiles
deviates from a Gaussian (see Sect.~5.1 on how such a case can be dealt with).

\begin{table}[!t]
\begin{center}
\caption{Values of ${\cal{W}}_{ul}$ for the spectral lines considered here.}
\begin{tabular}{cccc}
\hline\hline 
 $J_l$ & 	 $J_u$ & ${\cal{W}}_{ul}$ & Examples of spectral lines \\
\hline
 \multirow{2}{0.2cm}{$\displaystyle\frac{1}{2}$} &
 \multirow{2}{0.2cm}{$\displaystyle\frac{1}{2}$} & \multirow{2}{0.15cm}{0} 
 & \ion{Mg}{x} 624.941~{\AA}, \ion{O}{vi} 1037.61~{\AA} \\
   &   &  &  {\ion{H}{i}} 1215.6736~{\AA}\\
 \multirow{2}{0.2cm}{$\displaystyle\frac{1}{2}$} &
 \multirow{2}{0.2cm}{$\displaystyle\frac{3}{2}$} & \multirow{2}{0.35cm}{0.5} 
 & \ion{Mg}{x} 609.793~{\AA}, \ion{O}{vi} 1031.92~{\AA} \\
  &  &  & {\ion{H}{i}} 1215.6682~{\AA}\\
\hline
\end{tabular}
\label{wul_values}
\end{center}
\end{table}

 \begin{figure}[!th]
 \begin{center}
 {\resizebox{\hsize}{!}{\includegraphics{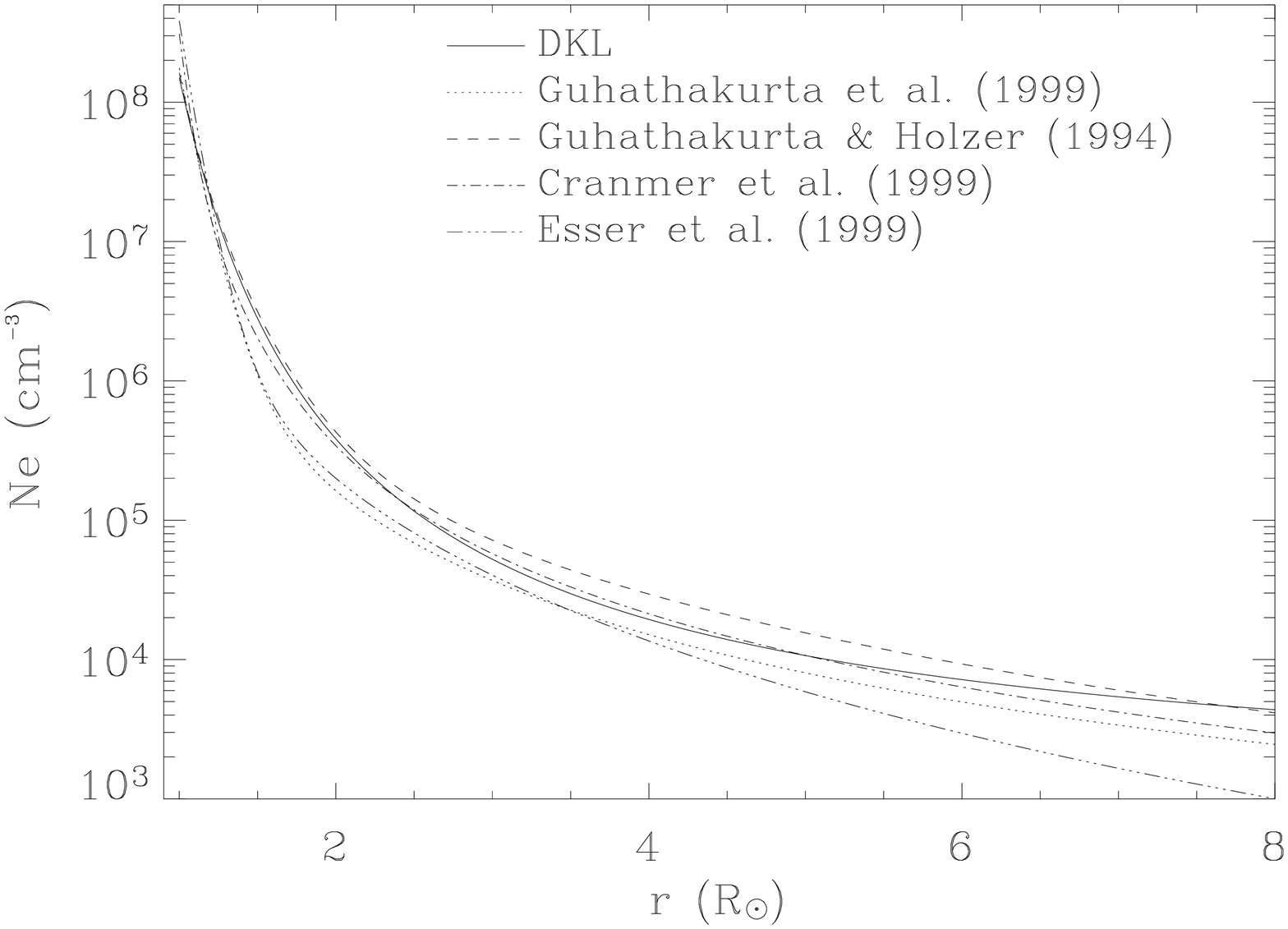}}}
 \caption{Electron density plotted as a function of the distance to Sun center for different
 empirical models (solid curve: DKL (Doyle et al. 1999a,b; Kohl et al. 1998; and Lamy et al. 1997)
 dotted: Guhathakurta et al. 1999; Dashes: Guhathakurta \& Holzer 1994; Dot-dashes: Cranmer et al.
 1999c; Triple-dot-dashes: Esser et al. 1999).
 \label{Elecdens_Density}}
 \end{center}
 \end{figure}

According to Eq.~(\ref{Gen_Emis_Formula}), the profiles created by collisions
and by resonant scattering at a given point along the LOS all have Gaussian
shapes. However, there are significant differences between the two components.
The main differences are: \\
 1) The intensity of the collisional component is
proportional to the square of the electron density, where as the radiative
component is only proportional to the density. Thus, the collisional component
suffers the radial density drop more than the radiative one. \\
 2) The
radiative component is affected by the outflow speed of the scattering ions
through the Doppler dimming effect, while the collisional component is not.
Thus, the contributions to the collisional component from sections along the LOS
supporting large outflow speeds are more important than the contributions to the
radiative one, since the latter gets progressively out of resonance due to the
increase in outflow speed of the solar wind with the distance from Sun center.\\
 3) For a given point on the LOS, the Doppler shift of the collisional component
reflects the real LOS speed $u_Z$ of the scattering atoms/ions. The shift of the
radiative component, however, corresponds to only a fraction of this speed. This
point is very important for the broadening of the line profile. Note that in
frame of the complete redistribution approximation, there is no shift difference
between collisionally and radiatively excited profiles (see Sahal-Br\'echot \&
Raouafi 2005).

\section{Atmospheric parameters}

The realistic synthesis of spectral line profiles requires the modeling of all
the parameters entering the formation process of a given spectral line.
According to Eq.~(\ref{Gen_Emis_Formula}), this includes the magnetic field, the
solar wind outflow velocity, the density stratifications of electrons and of
atoms/ions, etc., at every point along the LOS. For the large scale magnetic
field of the corona and the solar wind velocity we adopt the same descriptions
as RS04, while for the density stratifications we consider a wider range of
models.

The coronal magnetic field is described by the model of Banaszkiewicz et al.
(1998; see Fig.~1 of RS04). Cranmer et al. (1999c) used the same model to
constrain the strength and superradial expansion of the magnetic field in the
corona. The solar wind outflow speed is obtained through the mass-flux
conservation equation
\begin{eqnarray} 
\begin{array}{c}
{\displaystyle V(r,\theta) = \frac{Ne(R_{\sun})}{Ne(r,\theta)} \, 
                      \frac{B(r,\theta)}{B(R_{\sun},\theta_{\sun})} \,
			    V(R_{\sun},\theta_{\sun}) },
\end{array}
\label{SW_OutFlowSpeed}
\end{eqnarray}
where $Ne(R_{\sun}),\,V(R_{\sun},\theta_{\sun})$ and $B(R_{\sun},\theta_{\sun})$
are the electron density, the outflow speed of the ions and the coronal magnetic
field at the base of the solar corona (solar surface), respectively.
$Ne(r,\theta),\;V(r,\theta)$ and $B(r,\theta)$ are the same quantities at
coordinates $(r,\theta)$, with $r>R_{\sun}$. The angles $\theta$ and
$\theta_{\sun}$ are related to each other by the requirement that
Eq.~(\ref{SW_OutFlowSpeed}) is valid along a field line. This also implies
that the velocity vector is parallel to the magnetic field vector (needed for
the calculation of the dimming rate of the radiative component and also for the
determination of the Doppler shift of the reemitted spectral line).  In order
to be consistent with Ulysses observations that the fast solar wind varies by
less than a few percent with latitude (Neugebauer et al. 1998; McComas et al.
2000; Zhang et al. 2002), we adopt $V(R_{\sun},\theta_{\sun})\propto
B(R_{\sun},\theta_{\sun})$ as boundary condition on the outflow speed on the
solar surface (see Table~\ref{VsuntoBsun}). We consider different values of
$V_{\sun}$ for the different species. We also consider different $V_{\sun}$ for
the different density stratification models in order to get as close as possible
to the observed values of widths, total intensities and intensity ratios.

\begin{table*}[!t]
\begin{center}
\caption{Proportionality coefficient, $\kappa$, of $V(R_{\sun},\theta_{\sun})$ to
$B(R_{\sun},\theta_{\sun})$
$\left(\displaystyle{{V(R_{\sun},\theta_{\sun})}=\kappa\times{B(R_{\sun},\theta_{\sun})}}\right)$
used for the calculations of the different spectral lines considered in the present paper and
corresponding to the different density models.}
\begin{tabular}{lccccc}
\hline\hline 
 & 	 \multirow{2}{0.65cm}{DKL} & Guhathakurta \& Holzer & Guhathakurta et al. & Cranmer et al. & Esser et al.  \\
 &  & (1994)    & (1999) & (1999c) & (1999) \\
\hline
 \ion{H}{i}  &  0.63  &  0.85 &  0.35 &  0.30 &  0.13 \\
 \ion{O}{vi} &  0.85  &  0.88 &  0.50 &  0.35 &  0.13 \\
 \ion{Mg}{x} &  0.85  &  0.88 &  0.50 &  0.35 &  0.13 \\
\hline
\end{tabular}
\label{VsuntoBsun}
\end{center}
\end{table*}

\begin{figure}[!h]
\begin{center}
{\resizebox{\hsize}{!}{\includegraphics{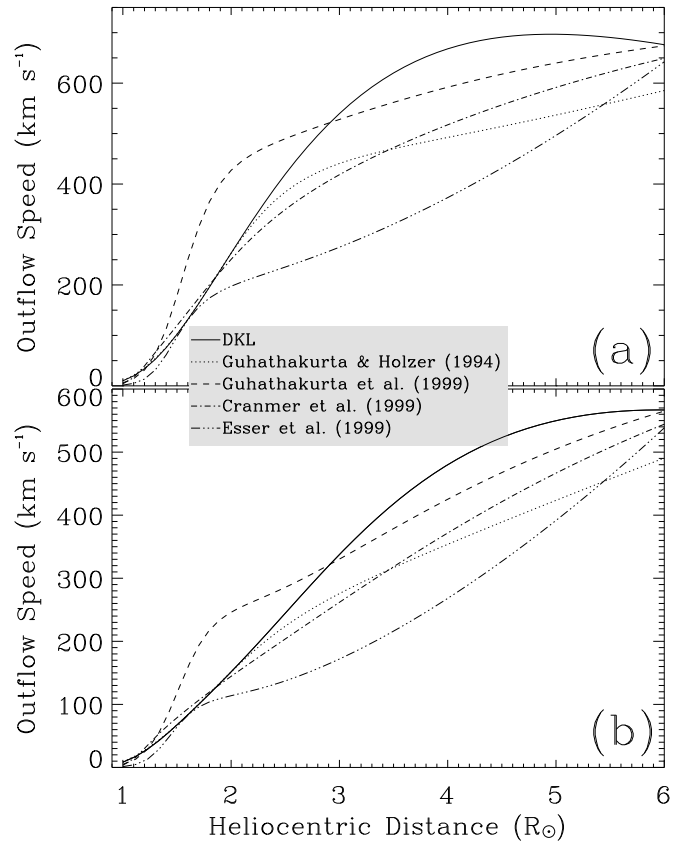}}}
\caption{Outflow speed of the \ion{O}{vi} ions along the polar axis and field line arising from 70
degree latitude, respectively, for the different density models considered. The Boundary condition on
the outflow speed of the coronal ions is chosen to be proportional to the solar surface magnetic field
strength (see text). In order to get as close as possible to the measured values, we adopt different
values of the proportionality coefficient of $V(R_{\sun},\theta_{\sun})$ with respect to
$B(R_{\sun},\theta_{\sun})$ for the different density models. The outflow speed of other species can be
obtained by multiplying the present ones by a constant.
\label{Outflows121204_NBCVSun_Pole_MFL70deg.eps}}
\end{center}
\end{figure}

RS04 have shown that the density stratification is a key quantity for
determining the off-limb profiles of collision-dominated lines, in particular
the line width. Here we consider a more complete set of empirical density
stratifications than in that paper. In addition to the electron density profiles
from (Doyle et al. 1999a,b; hereafter abbreviated as DKL) determined empirically
by combining the results obtained by three instruments on SOHO (SUMER: Doyle et
al. 1999a,b; UVCS: Kohl et al. 1999; LASCO: Lamy et al. 1997) and from
Guhathakurta \& Holzer (1994; SKYLAB) which were considered by RS04, we also
employ the models of Guhathakurta et al. (1999; SPARTAN), Cranmer et al. (1999c)
and Esser et al. (1999; SPARTAN and Mauna Loa; Fisher \& Guhathakurta 1995). The
absolute values of the density in all these models are not very different. They
are generally within the error bars of the coronal densities which are typically
20 - 30 \% of the measured values (the main exception is the Esser et al.
model). Except for the DKL model, the data for the other models has been fitted
by power series functions based on the fact that the density drops very fast
close to the Sun and follows an $\sim{r^{-2}}$ function sufficiently far away.
The description given by Doyle et al. for the DKL electron density
stratification (Eq.~(13) of RS04) assumes that the coronal gas is isothermal and
in hydrostatic equilibrium (see Guhathakurta \& Fisher 1995 \& 1998). 

The absolute value of the electron density directly affects the intensity of the
emitted line at any point along the LOS. However, the importance of the density
stratification resides therein that it also influences the LOS-integrated
profile indirectly through the solar wind speed, which is determined via
mass-flux conservation and thus depends on both the density stratification and
the magnetic field. Even somewhat different density stratifications give quite
different solar wind speeds and therefore very different line profiles. A more
thorough discussion is given in the following Sects. The effect of the density
stratification on the outflow speed and LOS velocity has been illustrated by
RS04; see their Fig.~5.

The solar wind speed profiles obtained for the combination of the chosen
magnetic structure and the density stratifications are plotted in
Fig.~\ref{Outflows121204_NBCVSun_Pole_MFL70deg.eps} for two extreme cases. In
Fig.~\ref{Outflows121204_NBCVSun_Pole_MFL70deg.eps}(a) the wind speed along the
polar axis is displayed, while
Fig.~\ref{Outflows121204_NBCVSun_Pole_MFL70deg.eps}(b) shows the speed along the
field line arising from 70 degree latitude on the solar surface. The LOS at
$3.5~R_{\sun}$ intersects this field line at $6~R_{\sun}$, which is the distance
to which we integrate the profiles.

The small decrease in the solar wind speed at high altitudes (solid line in
panel a) indicates some mismatch between the DKL model and the Banaszkiewicz et
al. magnetic structure right above the pole. This decrease does not affect our
conclusions since it occurs at altitudes above $4~R_{\sun}$, whereas the
furthest LOS we consider passes at $3.5~R_{\sun}$ above the pole.

The scaling factors of the solar wind speed at the solar surface are listed in
Table~\ref{VsuntoBsun}. Clearly although the scaling factors of the heavy
elements are not the same as for hydrogen, they differ by less than a factor of
1.5, with the hydrogen atoms moving more slowly.

In order to compute the total intensities of the lines emitted in the polar
coronal holes, we need the absolute density of each ion $N_{X^{n+}}$, a
parameter not considered by RS04. It can be written as follows
\begin{eqnarray}
\begin{array}{l}
{\displaystyle N_{X^{n+}}=\frac{N_{X^{n+}}}{N_{X}}\,
        \frac{N_{X}}{N_{H}}\,\frac{N_{H}}{N_{\rm{e}}}\,N_{\rm{e}}},
\label{IonizEquil}
\end{array}
\end{eqnarray}
where ${\displaystyle({N_{X^{n+}}}/{N_{X}}})$ is obtained from the ionization
balance of the $X$ species, ${\displaystyle({N_{X}}/{N_{H}}})$ is the abundance
of X relative to hydrogen and ${\displaystyle({N_{H}}/{N_{\rm{e}}}})$ is the
abundance of Hydrogen relative to electrons. Here we adopt a plasma with 10\% of
helium (the cosmic value) so that ${\displaystyle{N_{H}}/{N_{\rm{e}}}=0.83}$.
However, the coronal helium abundance could be considerably larger and could
vary significantly through the region considered here (see Joselyn \& Holzer
1978; Hansteen, Holzer \& Leer 1993, Hansteen, Leer \& Holzer 1994a,b \& 1997).
The considered fraction of neutral hydrogen is $2.667~10^{-7}$ (John Raymond,
private communication). ${\displaystyle{N_{O^{5+}}}/{N_{O}}}=4.78~10^{-3}$
(Nahar 1999) and the abundance of oxygen relative to hydrogen is taken to be 8.7
(Asplund et al. 2004).

\section{Effect of the integration along the LOS on the collisional and
radiative components}

The results that will be presented in the present Sect. are obtained by using
the DKL density stratification. The use of any of the other density models does
not change the conclusions.

UVCS samples three types of lines: the \ion{Mg}{x} lines at 609.793~{\AA}
and 624.941~{\AA} are only excited by electron collisions, while the Ly-$\alpha$
line is exclusively created by resonant scattering of the solar disk radiation.
The contribution of the collisional component is negligible (less than 1\%
according to Raymond et al. 1997). The {\ion{O}{vi}} lines at 1031.92~{\AA} and
1037.61~{\AA} are excited by the combination of collisions and scattering of the
transition region radiation.

Here we consider separately the collisional and radiative components of the
considered coronal spectral lines in order to illustrate better the effects of
the solar wind outflow velocity, density stratification and the integration
along the LOS on each of them. The behavior of a particular line depends on the
relative strengths of these two components.

\begin{figure}[!t]
\centering
{\resizebox{\hsize}{!}{\includegraphics{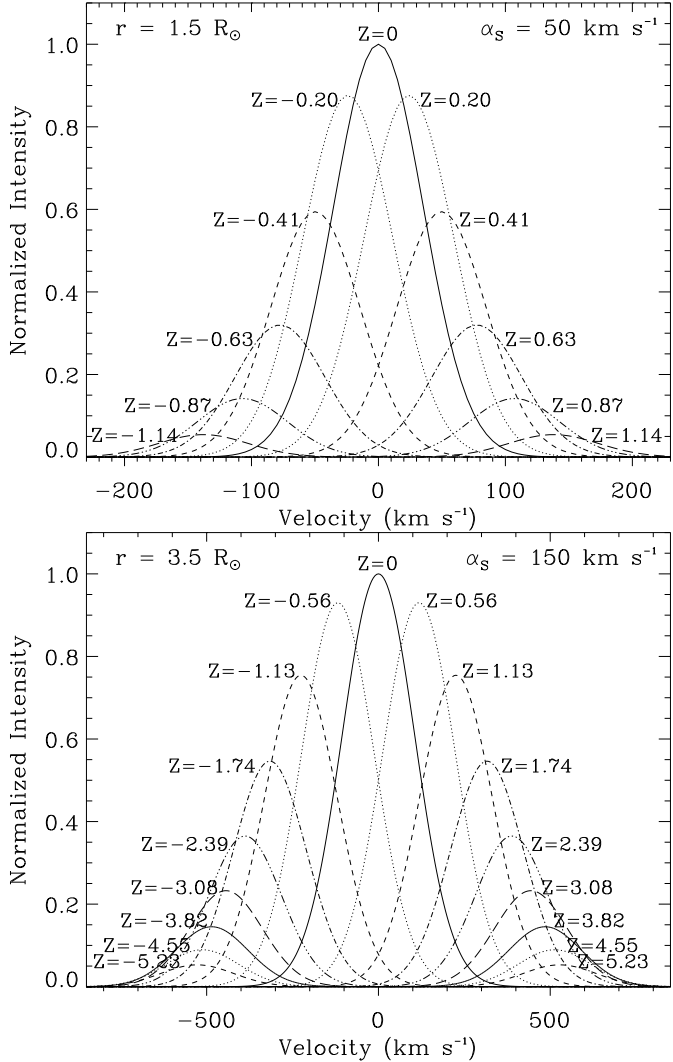}}}
\caption{Contributions to the final line profile emanating from different
sections of the  LOS (different values of $Z$) for a spectral line excited only
by electron collisions (namely a \ion{Mg}{x} doublet line) for LOSs coming to
within $0.5~R_{\sun}$ and $2.5~R_{\sun}$ of the solar surface (top and bottom
frames, respectively). The LOS integrated profile is the sum of all the
individual profiles emitted by the different sections along the LOS. The marked
$Z$ values are in units of $R_{\sun}$. Blue Doppler shifts correspond to
positive LOS speeds (toward the observer) that are obtained for positive values
of $Z$.
\label{LOS_Diff_Prof_MgX.eps}}
\end{figure}

\begin{figure}[!th]
\centering
{\resizebox{\hsize}{!}{\includegraphics{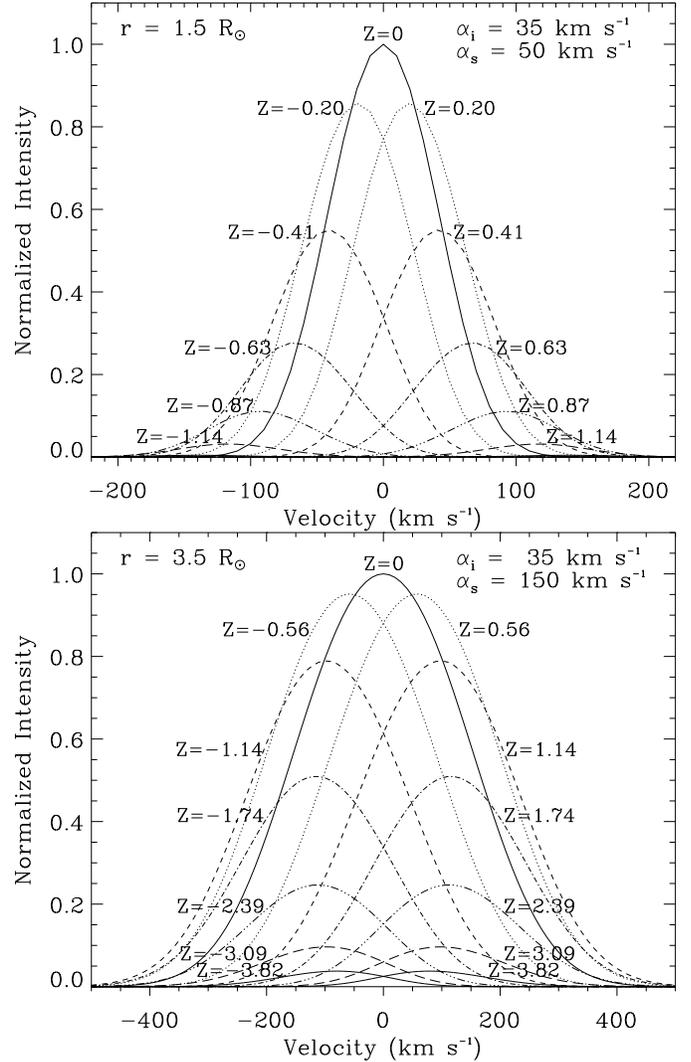}}}
\caption{Equivalent results as in Fig.\ref{LOS_Diff_Prof_MgX.eps} but for a purely radiatively
excited line.
\label{LOS_DiPr_RAD_R1.5_3.5_DOY_O6_270204_Pap1.eps}}
\end{figure}

\begin{figure*}[!t]
\centering
\includegraphics[width=0.48\textwidth]{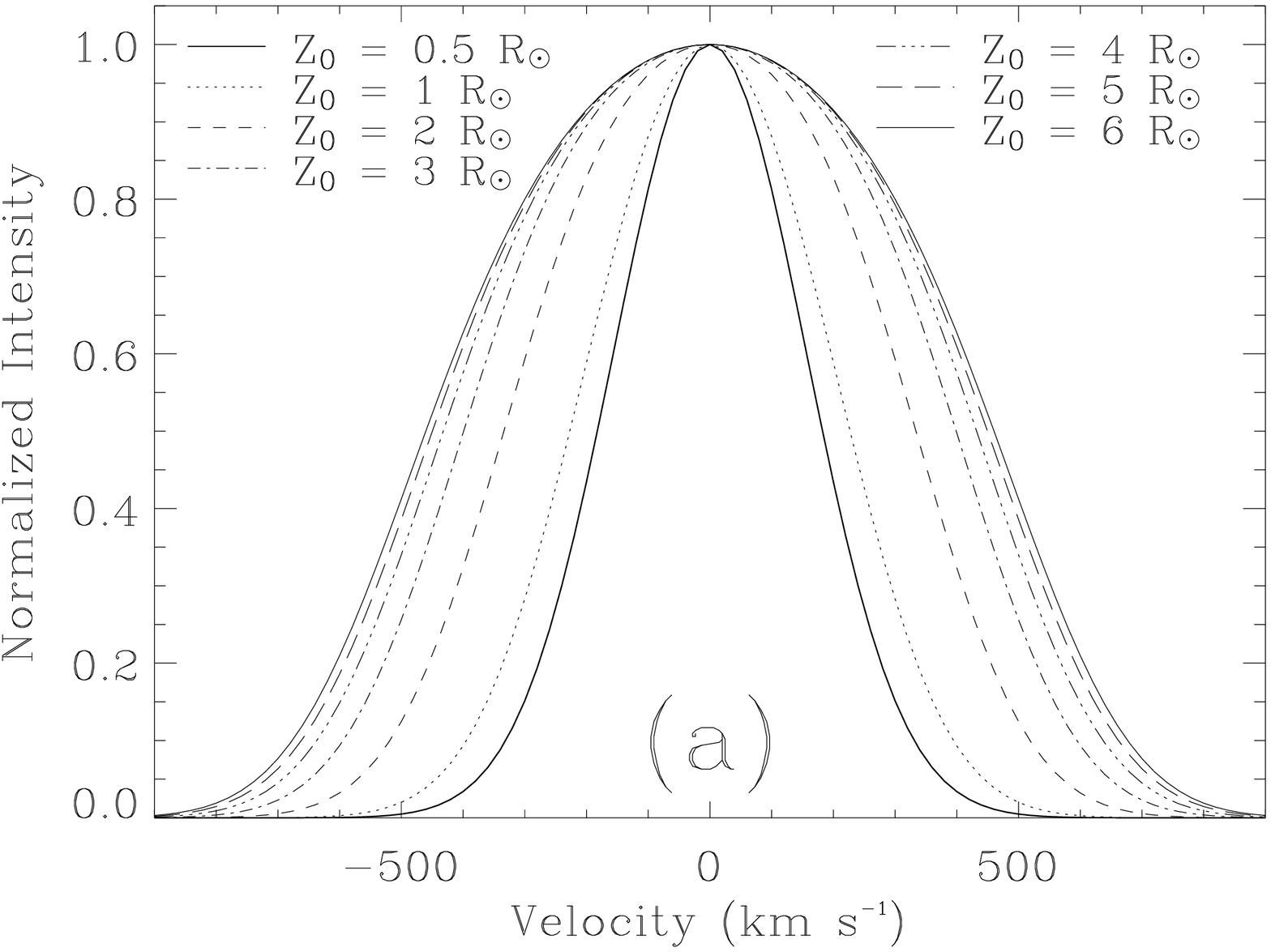}
\includegraphics[width=0.48\textwidth]{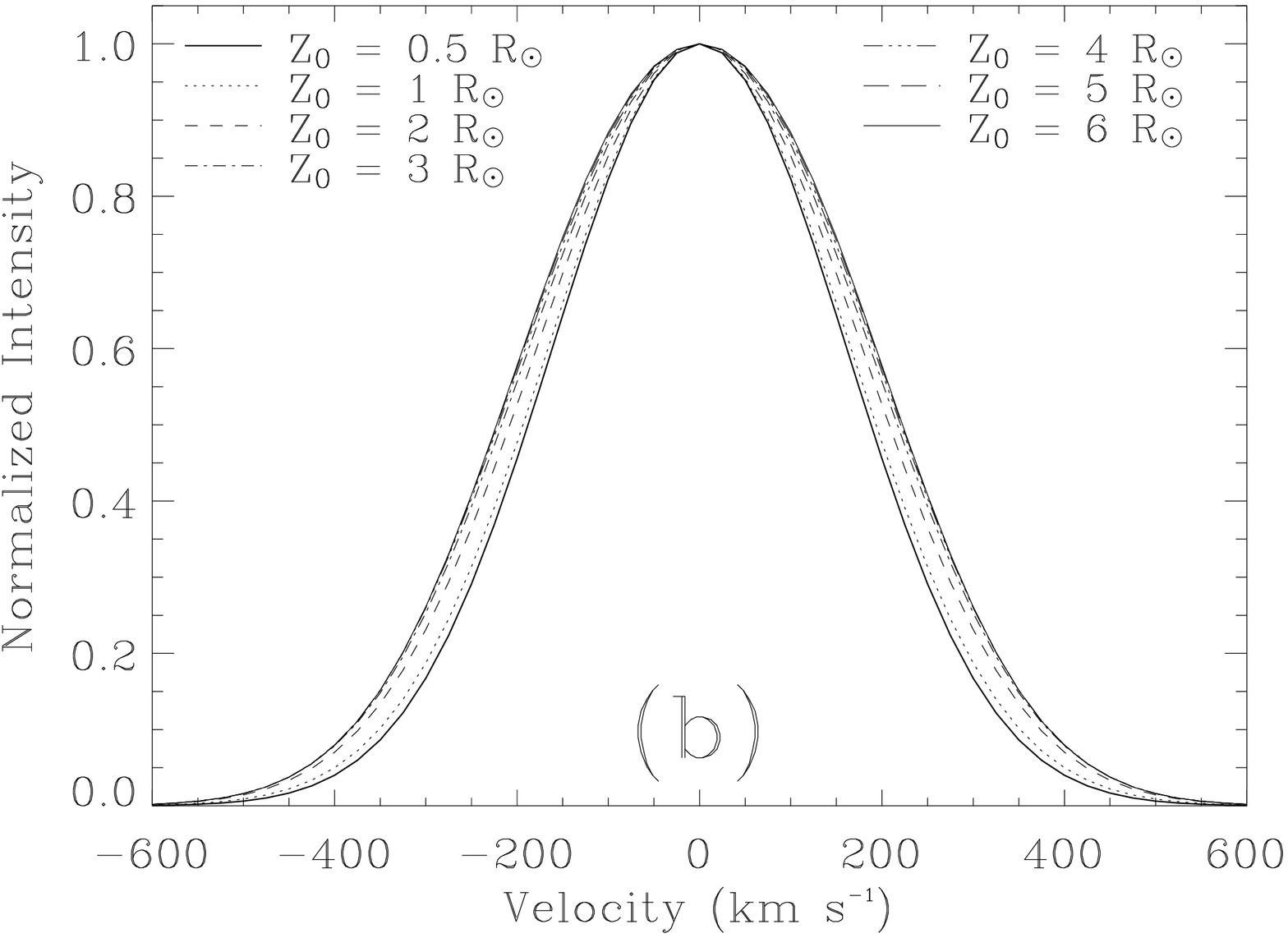}
\caption{Line profiles obtained by integrating over different distances between $-Z_0$ and $Z_0$
along the LOS for a spectral line excited only by electron collisions (a) and only by
radiation (b). These profiles are obtained for a point of closest approach of $3.5~R_{\sun}$
from Sun center and for $\alpha_s=200$ km~s$^{-1}$ ($\alpha_i=100$ km~s$^{-1}$ for the radiative
component). The effect of the LOS integration on the radiatively excited line is rather
small compared to the collisional line.
\label{Prof_MgX_RayInt_R3.5.eps}}
\end{figure*}

Figs.~\ref{LOS_Diff_Prof_MgX.eps} and
\ref{LOS_DiPr_RAD_R1.5_3.5_DOY_O6_270204_Pap1.eps} display the profiles emitted
from different sections along the LOS for two different projected heliocentric
distances 1.5 and $3.5~R_{\sun}$ (i.e. different rays) for spectral lines
created solely by electron collisions and purely from resonant scattering of the
solar disk radiation, respectively. For the sake of clarity, we used
constant $\alpha_s$ to illustrate the effect of the LOS integration on the
profiles in Figs.~\ref{LOS_Diff_Prof_MgX.eps},
\ref{LOS_DiPr_RAD_R1.5_3.5_DOY_O6_270204_Pap1.eps} \&
\ref{Prof_MgX_RayInt_R3.5.eps}. When comparing with observed profiles in Sect.
5, however, we only consider computations with height-dependent $\alpha_s$.

At low altitudes ($<2~R_{\sun}$) the main contribution is given by the central
part of the LOS (i.e. near the polar axis). This is due to the fast drop of the
densities as a function of the heliocentric distance
(Fig.\ref{Elecdens_Density}). However, when further from the solar disk,
significant contributions are obtained from larger sections (larger $|Z|$) along
the LOS, due to the relatively slow decrease of the densities far from the solar
limb ( Fig.~\ref{Elecdens_Density}). Doppler dimming, which is largest at
greater heights, concentrates the contribution of the radiative component to the
observed line profile toward small $|Z|$ (compare the bottom panels of
Figs.~\ref{LOS_Diff_Prof_MgX.eps} \&
\ref{LOS_DiPr_RAD_R1.5_3.5_DOY_O6_270204_Pap1.eps}). Another striking difference
is seen in the Doppler shift behavior of the two components. The Doppler shift
of the collisional profiles increases when moving away from the polar axis along
the LOS (Fig.~\ref{LOS_Diff_Prof_MgX.eps}). The Doppler sift of the radiative
component also increases at small $|Z|$, but ever more slowly at larger $|Z|$.
Depending on the profile of the LOS velocity, the line shift can reach a
limiting value characteristic of that LOS, or can even start to decrease again
at greater $|Z|$ (Fig.~\ref{LOS_DiPr_RAD_R1.5_3.5_DOY_O6_270204_Pap1.eps}).

The behavior seen in Figs.~\ref{LOS_Diff_Prof_MgX.eps} \&
\ref{LOS_DiPr_RAD_R1.5_3.5_DOY_O6_270204_Pap1.eps} can be understood as follows.
At a given $Z$, the Doppler shift of the collisional component is equal to the
LOS speed $u_Z$ of the scattering atoms/ions (see first term of the RHS of
Eq.~(\ref{Gen_Emis_Formula})). However, for the radiative component the Doppler
shift is given by
${\displaystyle(u_Z-\alpha_{si}\,n_Z\,\textbf{\textit{u}}\cdot\textbf{\textit{n}})}$.
For small $|Z|$ (close to the pole where the field lines are only slightly
inclined with respect to the polar axis), $n_Z$ is very small and the Doppler
shift is approximately equal to $u_Z$, so that collisional and radiative line
profiles arising at small $|Z|$ are similar. However, further away from the
polar axis, $n_Z$ increases (in absolute value) so that, depending on $n_Z$ and
$\alpha_{si}$ $(=\alpha_s^2/(\alpha_i^2+\alpha_s^2))$, the expression
${\displaystyle(u_Z-\alpha_{si}\,n_Z\,\textbf{\textit{u}}\cdot\textbf{\textit{n}})}$
either saturates or even decreases again ($\alpha_i$ is the Doppler width of the
incident line profile). This effect is more marked at higher altitudes where the
contribution from LOS sections with large $|Z|$ are important (see bottom panel
of Fig.~\ref{LOS_DiPr_RAD_R1.5_3.5_DOY_O6_270204_Pap1.eps}).

In Fig.~\ref{Prof_MgX_RayInt_R3.5.eps} profiles for a purely collisionally (a)
and purely radiatively (b) excited line are displayed. The plotted lines are
obtained by integrating out to different distances along the LOS for a projected
heliocentric distance of $3.5~R_{\sun}$. Profiles of collisional lines turn out
to be sensitive to the atmospheric parameters, in particular LOS velocity, out
to $|Z|=6~R_{\sun}$. The profile shape of the radiatively excited line is hardly
altered by LOS integration, although the total intensity is significantly
affected (not visible in Fig.~\ref{Prof_MgX_RayInt_R3.5.eps} due to the
normalization).

\section{Application to coronal lines}

Among the spectral lines observed by UVCS in polar coronal holes three sets are
of particular interest. The first is the most intense line emitted in the solar
corona, Ly-$\alpha$. This line is produced by the resonant scattering of the
radiation emitted by the solar disk. The few electron collisions are negligible
for this line (Raymond et al. 1997). The second group of lines is the
{\ion{O}{vi}} doublet. These lines are excited by both, the radiation coming
from the underlying chromosphere-corona transition zone and by isotropic
electron collisions. The third class is composed of the {\ion{Mg}{x}} doublet,
whose members are emitted following excitations produced almost exclusively by
electron collisions (Wilhelm et al. 2004).

The dependence of $\alpha_s$ on $r$ and $\theta$ is computed consistently. We
assume that $\alpha_s$ depends only on $r$, so that its value changes along the
LOS. In the next subsections, we describe calculations of the profiles of these
three sets of lines and their comparison with the observed ones.

\subsection{The Ly-$\alpha$ line of hydrogen}

\begin{figure}[!t]
\centering
{\resizebox{\hsize}{!}{\includegraphics{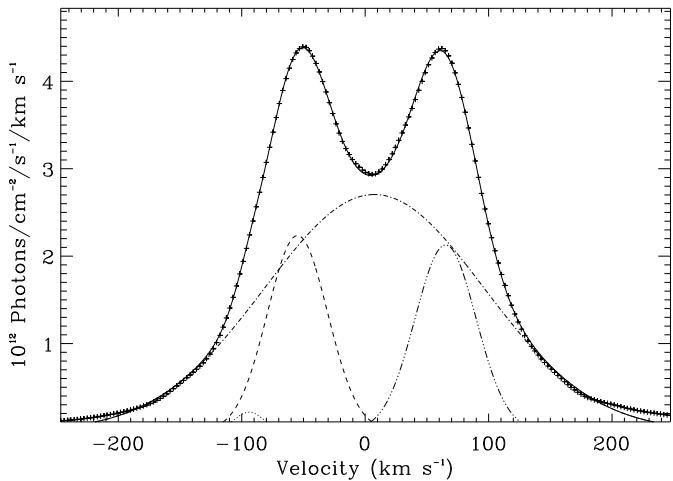}}}
\caption{Disk emission (+ signs) of Ly-$\alpha$ during the minimum of solar
activity (Lemaire et al. 1998). The solid curve is a four-Gaussian fit to the data. The
individual Gaussians contributing to the fit are given by the dotted, dashed, dot-dashed and
triple-dot-dashed lines.
\label{Lyman_Alpha_4G_Fit.eps}}
\end{figure}

To calculate the coronal emission H~Ly-$\alpha$, it is necessary to know the
shape of the line profile coming from the solar disk
(Fig.~\ref{Lyman_Alpha_4G_Fit.eps}). A four Gaussian fit plus a constant
background is applied to the complex shape of the observed profile. Numerically,
we assume that coronal hydrogen atoms are illuminated by the photons of four
incident Gaussian profiles emitted by the solar disk, with of course different
parameters, as listed in Table~\ref{H_I_table}. We do not consider any
center-to-limb variation in the intensity emitted by the solar disk in
Ly-$\alpha$ ($f(\Omega,\theta)=1$), which is in accordance with the filter
images of Bonnet et al. (1980).

\begin{figure}[!th]
\centering
{\resizebox{\hsize}{!}{\includegraphics{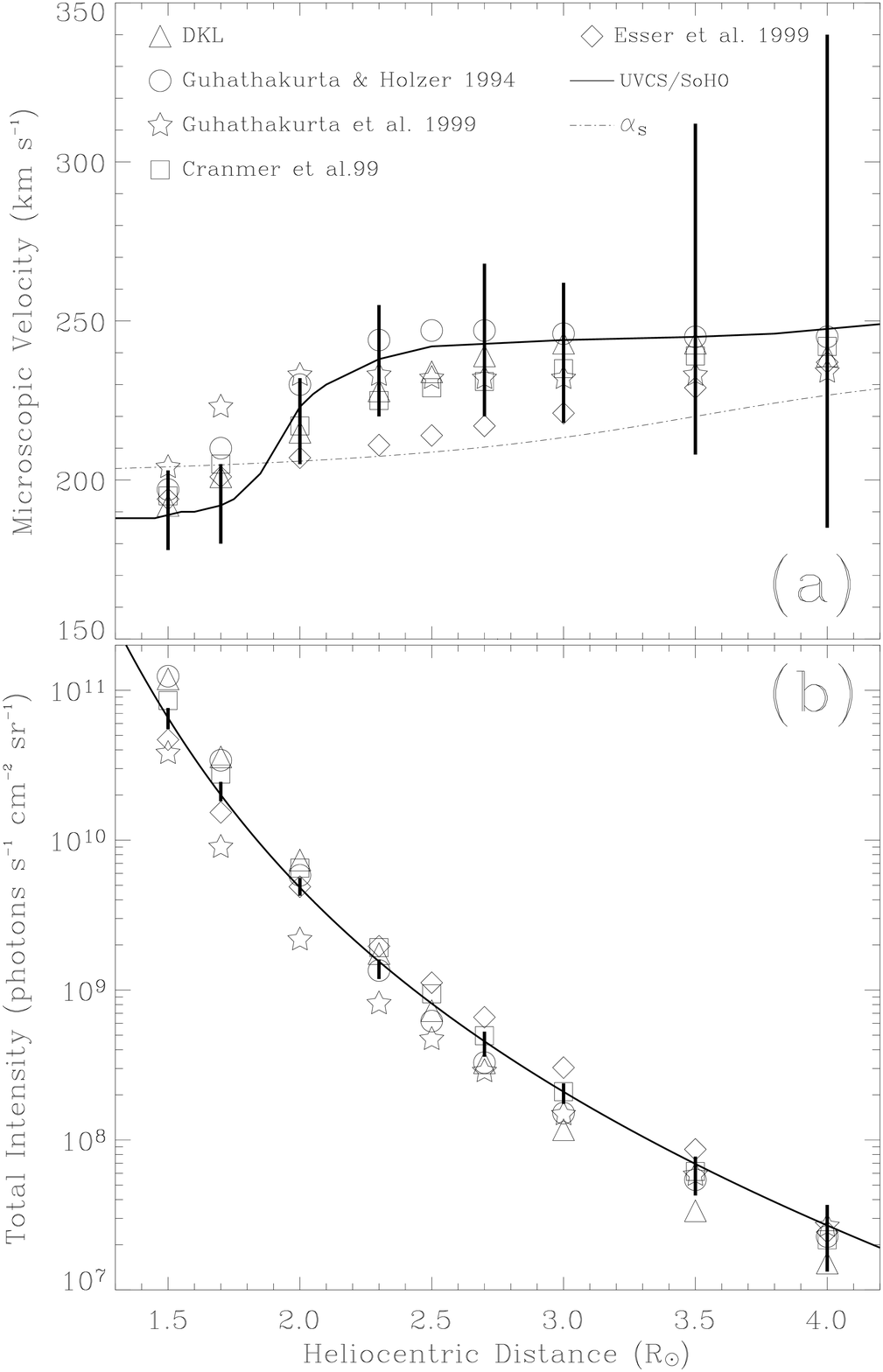}}}
\caption{Widths (a) and total intensities (b) of the LOS-integrated Ly-$\alpha$
line profile as a function of the projected heliocentric distance obtained for
the different density stratification models presented in
Fig.~\ref{Elecdens_Density}. The correspondence symbols to density models are
given in the top panel. The solid lines is the best fits to the observations
recorded by UVCS and the dot-dashed line (top panel) is the turbulence velocity
distribution of the hydrogen atoms. The vertical lines are the statistic error
bars of the measurements.  \label{Width_LyAlpha.eps}}
\end{figure}

\begin{table}
\begin{center}
\caption{Parameters of the different Gaussians used to fit the Ly-$\alpha$ line. The
Gaussians' centers are given in (km~s$^{-1}$) with respect to the line center as
obtained from the data (+ signs in Fig.~\ref{Lyman_Alpha_4G_Fit.eps}). The amplitudes are in
($10^{12}$~photons/cm$^{2}$/s/(km~s$^{-1}$) and the widths are in (km~s$^{-1}$).}
\begin{tabular}{lrrr}
\hline\hline 
Gaussian & Amplitude                              & Position      & Width \\
\hline
Dotted            & 0.22   & -94.01 & 16.81 \\
Dashed            & 2.24   & -54.79 & 34.39 \\
Dot-dashed        & 2.70   &   7.41 & 125.28 \\
Triple-dot-dashed & 2.13   &  65.13 & 34.91 \\
\hline
\end{tabular}
\label{H_I_table}
\end{center}
\end{table}

A single Gaussian is sufficient to fit the calculated off-limb Ly-$\alpha$
profiles with good accuracy. Fig.~\ref{Width_LyAlpha.eps}(a) displays the
e-folding (Doppler) widths of the LOS integrated Ly-$\alpha$ profiles as a
function of the projected heliocentric distance on the plane of the sky (the
widths of the best-fit Gaussian are plotted). The values of $\alpha_s$ used to
reproduce these profiles are also given in the same Figure (dot-dashed line)
together with the best fit to the observational data (solid line) by Cranmer et
al. (1999a) (see also Kohl et al. 1998).

The computed Ly-$\alpha$ profiles are relatively in-sensitive to the details of
the electron density stratification. For most of the density stratifications
given in Fig.~\ref{Elecdens_Density}, the observed line width of Ly-$\alpha$ in
polar coronal holes at different heights above the solar limb are comparable to
the calculated ones in the presence of an isotropic velocity distribution and a
gradual rise in $\alpha_s$ with projected distance to the limb. The displayed
results are all located in the 1-$\sigma$ error bars of measured widths except
for the widths obtained at low altitudes by using densities given by
Guhathakurta et al. (1999) or between 2.3 and $2.7~R_{\sun}$ based on the Esser
et al. densities. The reasons for these departures are visible in
Fig.~\ref{Outflows121204_NBCVSun_Pole_MFL70deg.eps}; note the high/low outflow
speeds obtained through these density models at these altitudes. Note that the
outflow speed of the hydrogen atoms at the solar surface is a free
parameter. By lowering the outflow speed at $1~R_{\sun}$ (solar surface)
for Guhathakurta et al. (1999) we can obtain a better correspondence with the
measured widths a low altitudes. However, this  occurs only at the cost of lower
widths also at higher altitudes, so that the overall correspondence with the
data is not improved.

Fig.~\ref{Width_LyAlpha.eps}(a) shows how the contribution to the line width
from the small-scale velocity distribution, given by $\alpha_s$, which dominates
at small $r$ is replaced by contributions from the outflowing solar wind at
larger $r$, as the line width significantly exceeds the $\alpha_s$ value.

Fig.~\ref{Width_LyAlpha.eps}(a) also demonstrates that a more rapid drop in
density with height leads to broader line profiles (compare profiles resulting
from the DKL stratification with that of Guhathakurta et al. 1999). This means
that of the two oppositely directed effects introduced by a steeper density
gradient, a more rapidly increasing wind speed and a smaller contribution to the
measured line profile from large $|Z|$ values, the former dominates.

Fig.~\ref{Width_LyAlpha.eps}(b) displays the total intensity of Ly-$\alpha$ as a
function of the projected heliocentric distance for the different density
stratifications considered in the present paper. The best fit to the observed
intensities (solid curve) and the error bars of the measurements (vertical bars)
are also plotted (see Cranmer et al. 1999a). All the density models considered
give total intensities that are close to the observed ones (within a factor of
$2-3$).

\subsection{{\ion{O}{vi}} doublet}

We compute the intensity profiles of the lines of the \ion{O}{vi} doublet at
1031.92~{\AA} $\left(2s\;^2{\rm{S}}_{1/2} - 2p\;^2{\rm{P}}_{3/2}\right)$ and
1037.61~{\AA} $\left(2s\;^2{\rm{S}}_{1/2} - 2p\;^2{\rm{P}}_{1/2}\right)$. In the
solar corona, the \ion{O}{vi} ions interact with electrons (with isotropic
velocity distribution) and with the radiation coming from the underlying
transition region. The effects of the ions' motion are taken into account via
the Doppler dimming, the Doppler shift and the optical pumping of the
\ion{O}{vi} 1037.61~{\AA} line by the chromospheric \ion{C}{ii} doublet
(1036.3367~{\AA} \& 1037.0182~{\AA}). This last effect is important above
$\sim2~R_{\sun}$ where the solar wind outflow speed reaches values that enable
optical pumping to occur (the exact radius at which this occurs depends on the
adopted solar wind profile). The incident solar disk spectrum considered for the
calculations of the coronal \ion{O}{vi} lines is shown in Fig.~2 of RS04. All
the line profiles are assumed to be Gaussians. In the present computations, the
radiances of the individual members of the \ion{O}{vi} doublet on the disk
spectrum are equated to 305 and 152.5 erg~cm$^{-2}$~s$^{-1}$~sr$^{-1}$,
respectively. Both \ion{O}{vi} incident lines are assumed to have the same
widths, corresponding to 35 km~s$^{-1}$. We consider also two identical profiles
for the \ion{C}{ii} doublet with widths of 25 km~s$^{-1}$ and radiances of 52
erg~cm$^{-2}$~s$^{-1}$~sr$^{-1}$ each. We use the limb-brightening measured by
Raouafi et al. (2002) for the \ion{O}{vi} line and no limb-brightening for the
\ion{C}{ii} lines (see Warren et al. 1998). For comparison, Table~1 of RS04
presents different values of these parameters measured from different
observations.

\begin{figure}[!t]
\centering
{\resizebox{\hsize}{!}{\includegraphics{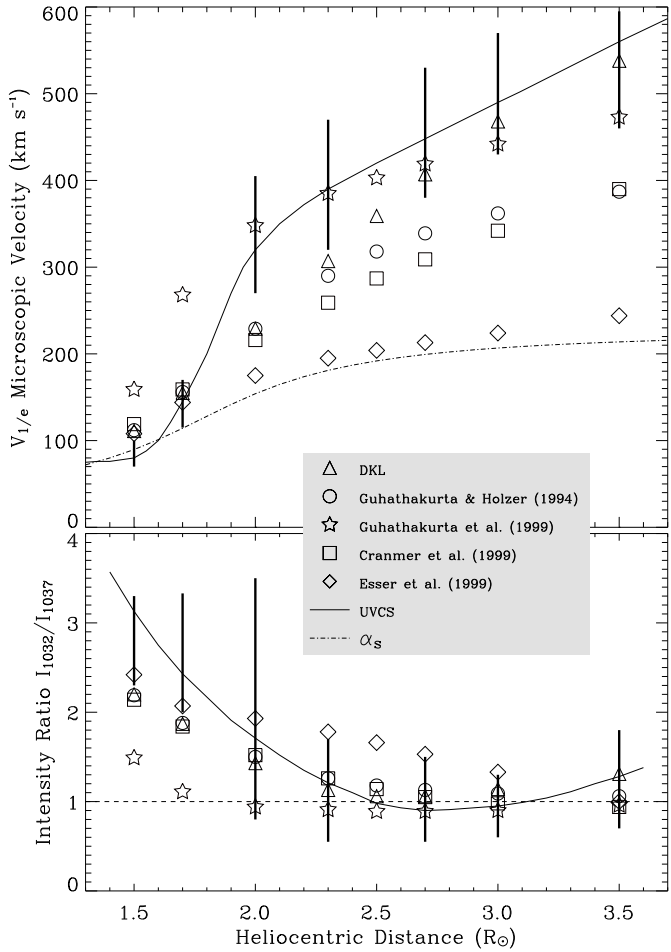}}}
\caption{The symbols represent the variation of the width (top panel) and intensity ratio
(bottom panel) of the LOS integrated synthetic line profiles
of \ion{O}{vi} as a function of the heliocentric distance. The correspondence between symbols and
models is indicated in the grey frame. The dot-dashed curve displays the used values of
$\alpha_s$. The observed dependence is given by the solid curves (best fits to the observations;
see Cranmer et al. 1999c), with the vertical lines being error bars.
\label{Line_Width_Ratio_New_DOYGUH_Paper.eps}}
\end{figure}

\begin{figure}[!t]
\centering
{\resizebox{\hsize}{!}{\includegraphics{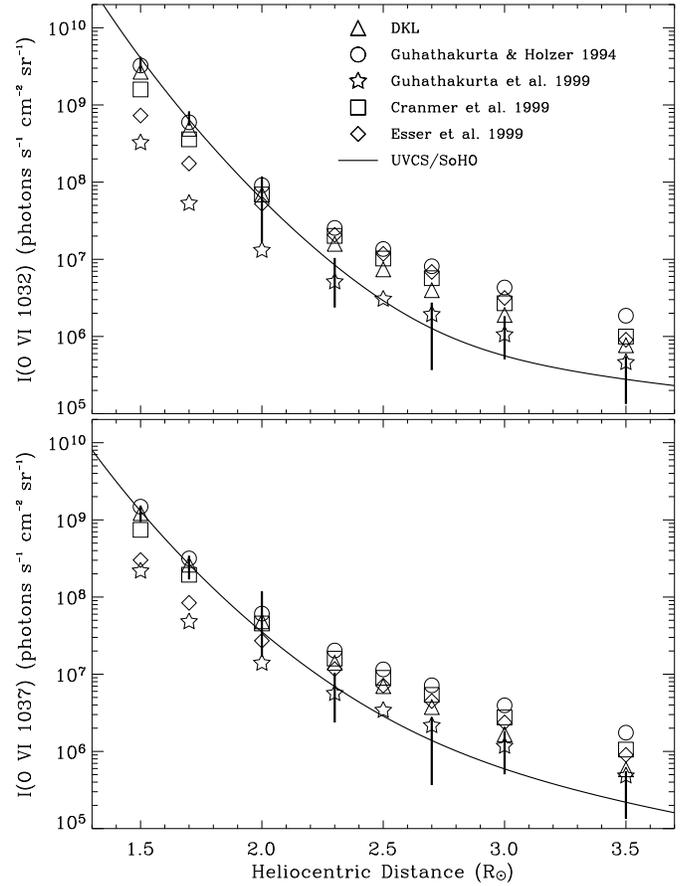}}}
\caption{Total intensities of the \ion{O}{vi} lines as a function of the projected heliocentric
distance. The correspondence to the different density stratifications is given by the symbols in
the top panel. The solid curves are the best fits to the observed intensities by UVCS and the
vertical bars represent estimates of the accuracy of the observations.
\label{TI3237_191004_NBCVSun.eps}}
\end{figure}

\begin{figure}[!t]
\centering
{\resizebox{\hsize}{!}{\includegraphics{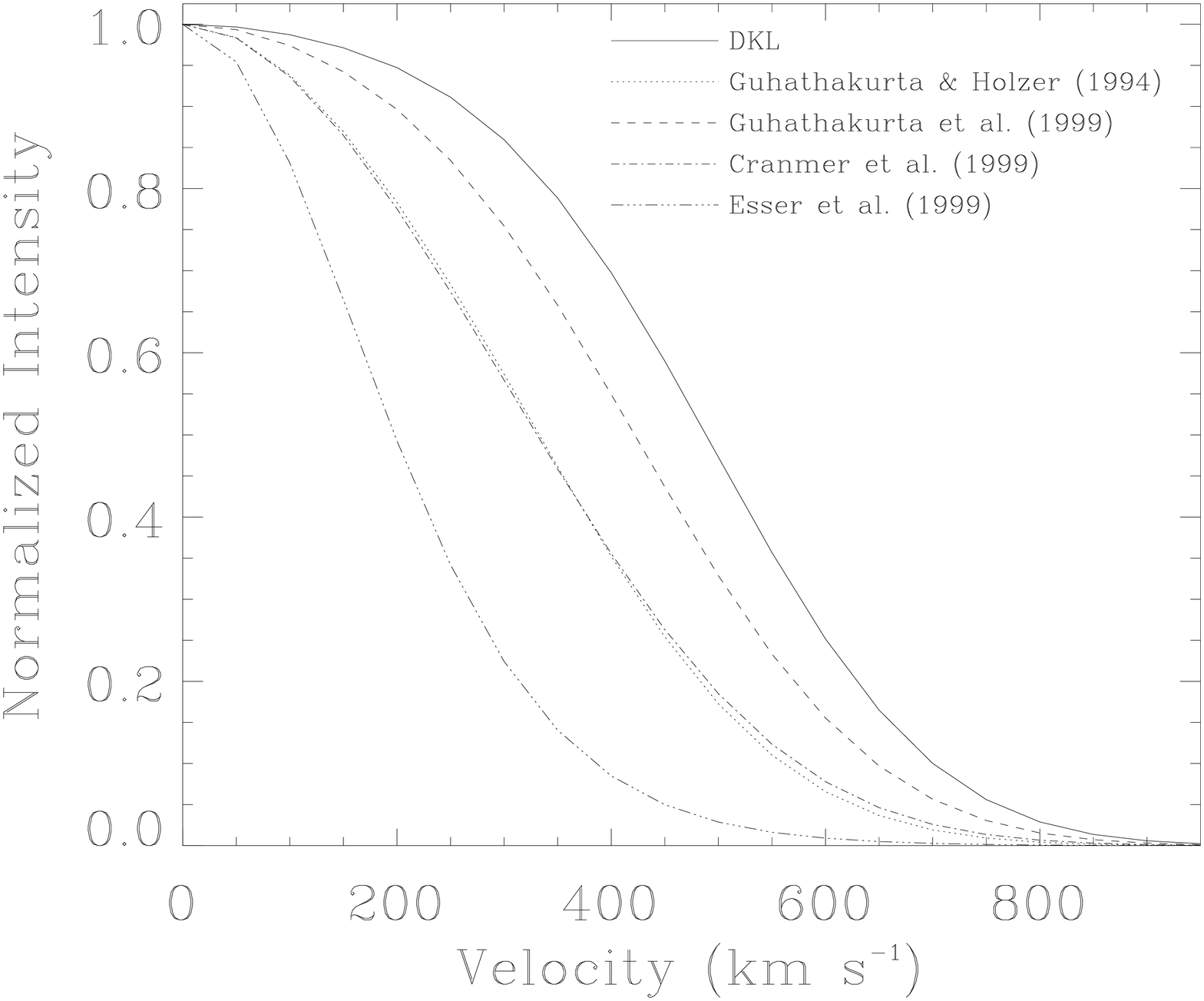}}}
\caption{LOS integrated profiles of the \ion{O}{vi}~1031.92~{\AA} line obtained at
a projected heliocentric distance of $3.5~R_{\sun}$ for the different electron density
stratifications considered in the present paper. The obtained line profiles are all approximatively
Gaussian shaped, although they differ strongly in width.
\label{Profiles_OVI_241104.eps}}
\end{figure}

The top panel of Fig.~\ref{Line_Width_Ratio_New_DOYGUH_Paper.eps} displays the
widths of the \ion{O}{vi} 1031.92~{\AA} line calculated for rays (LOSs)
corresponding to different projected heliocentric distances and for different
density stratification models. These widths are obtained by applying a Gaussian
fit to each calculated profile. All profiles are well represented by a Gaussian.
The profiles at the furthest considered ray ($3.5~R_{\sun}$) show the strongest
departure from a Gaussian shape. They are plotted for all considered density
models in Fig.~\ref{Profiles_OVI_241104.eps}. The solid line represents the best
fit to the UVCS data (Cranmer et al. 1999a). Note that no anisotropy in the
kinetic temperature of the emitting ions is considered.

At small heights ($<2~R_{\sun}$) the widths of the calculated profile are
comparable for most of the density models except for the one by Guhathakurta et
al. (1999). For this model the boundary condition on the solar wind speed at the
solar surface is chosen to fit the widths at high latitudes. This gives higher
solar wind speeds already at $1.5~R_{\sun}$ and explains the broader profiles
resulting from this model at low altitudes. At larger heights $(>2.0~R_{\sun})$
the line widths are very sensitive to the details  of the electron density
stratification. This can be easily seen by the difference in the width of the
different profiles obtained through slightly different density stratification
models. The model by DKL gives line widths comparable to the ones obtained from
the data, except between 2 and $2.7~R_{\sun}$, where the obtained widths are
slightly smaller than the observed ones.

The bottom panel of Fig.~\ref{Line_Width_Ratio_New_DOYGUH_Paper.eps} displays
the ratio of total intensities of the \ion{O}{vi} doublet lines
$({\cal{I}}_{1032}/{\cal{I}}_{1037})$ as a function a the projected heliocentric
distance. This ratio exhibits a marked dependence on radial distance being well
over 2 close to the Sun, then dropping rapidly (except for the density models by
Esser et al. 1999, which produces larger ratios compared to the ones given by
the other density models and which do not fit the observations in particular at
distances between 2.3 and  $3~R_{\sun}$). All the other models lead to a minimum
in the ratio at $r\approx2.5 - 3~R_{\sun}$, which then increases again at larger
$r$ (a number of the considered models exhibit a slightly different behavior,
showing a slight decrease in the ratio out to $r=3.5~R_{\sun}$). Generally, most
of the calculated intensity ratios (in particular those obtained from the DKL
density stratification) are within the error bars of the ones observed by UVCS.

Fig.~\ref{TI3237_191004_NBCVSun.eps} displays the computed total intensities of
the \ion{O}{vi} lines as a function of the projected heliocentric distance. The
density model of DKL, Cranmer et al., Guhathakurta \& Holzer, and for
$r\ge2~R_{\sun}$, Esser et al. give comparable intensities of the \ion{O}{vi}
doublet that agree relatively reasonably with respect to the observations. All
models produce a slightly less steep drop in intensity with altitude than
suggested by the observations, however. The density models by Guhathakurta et
al. (1999) and Esser et al. (1999) give low intensities at low altitudes. This
is due the fast drop of the electron density at low altitudes for these two
models.

\subsection{{\ion{Mg}{x}} doublet}

\begin{figure}[!h]
\centering
{\resizebox{\hsize}{!}{\includegraphics{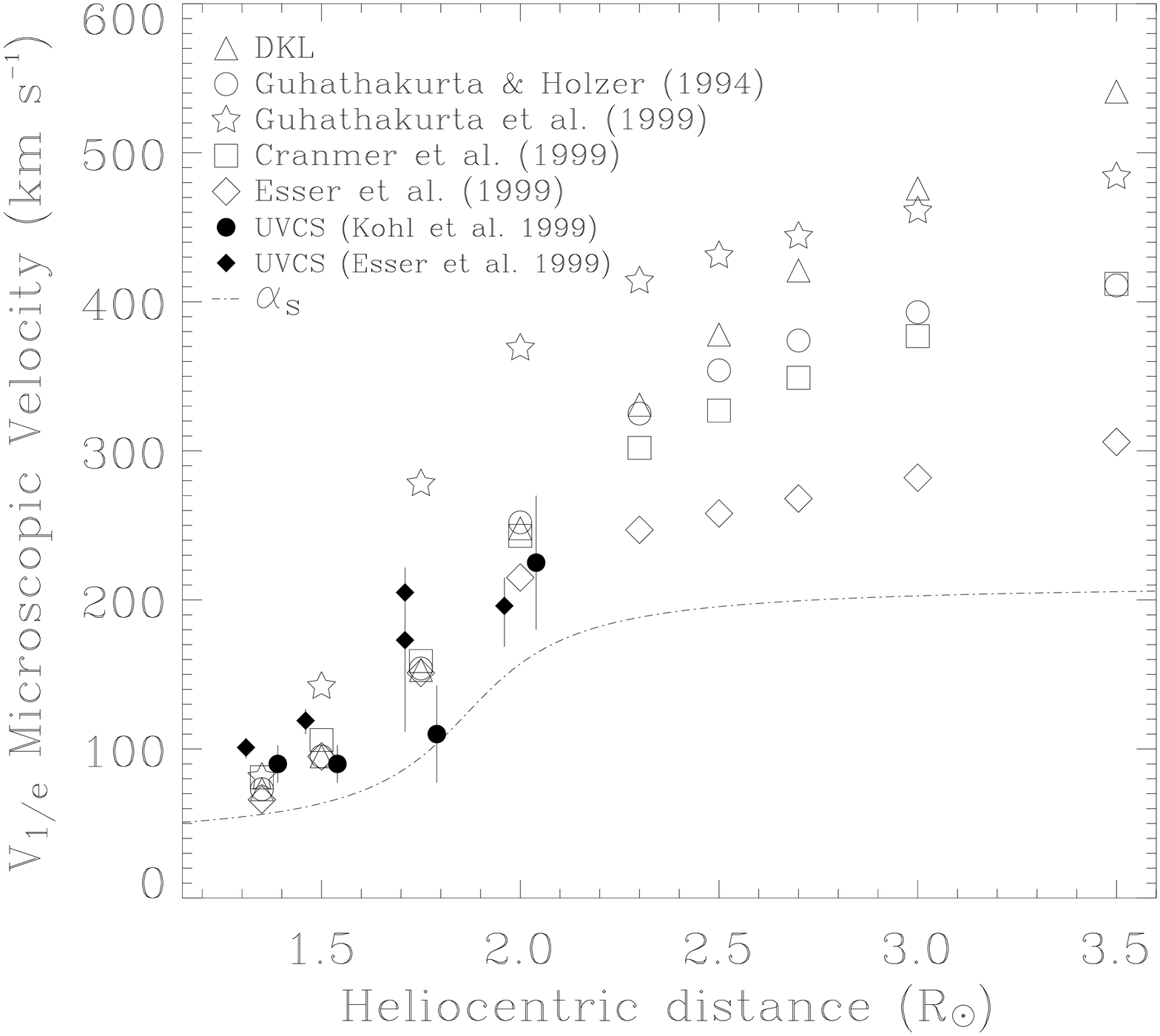}}}
\caption{Widths of the LOS integrated line profiles of the {\ion{Mg}{x}} doublet plotted
as a function of the projected heliocentric distance for the different density stratification
models considered in the present paper. The dot-dashed line gives $\alpha_s$ as a function of the
heliocentric distance. Also plotted are the measured widths of the {\ion{Mg}{x}} lines together
with the error bars. For the sake of clarity they are plotted slightly to the right (full circles;
data of Kohl et al. 1999) and to the left (full diamonds; data of Esser et al. 1999) from their
respective x-coordinates.
\label{Line_Width_MgX_New271103_AlpsVar.eps}}  
\end{figure}

The coronal {\ion{Mg}{x}} ion emits a doublet at 609.793~{\AA} and 624.941~{\AA}
that corresponds to the atomic transitions $2s\;^2{\rm{S}}_{1/2} -
2p\;^2{\rm{P}}_{3/2}$ and $2s\;^2{\rm{S}}_{1/2} - 2p\;^2{\rm{P}}_{1/2}$,
respectively. They are formed in the solar corona almost exclusively by electron
collisions (the disk emission is almost non existent; see Curdt et al. 2001).
Due to the weakness of the coronal emission in these lines, UVCS was only able
to record data in the polar coronal holes up to a height of 2~$R_{\sun}$ from
Sun center. The members of the doublet have almost identical widths, so that a
single value suffices to describe both of them. The widths of the observed
profiles are displayed in Fig.~\ref{Line_Width_MgX_New271103_AlpsVar.eps}
together with their error bars. Note the small, but systematic difference
between the values determined by Esser et al. (1999) and Kohl et al. (1999). The
widths of the calculated profiles and the values of microscopic velocities
$\alpha_s$ used to reproduce these profiles are plotted in the same Figure. The
obtained synthetic profile widths are comparable with the observed widths within
the accuracy of the data, in particular if we consider the scatter of the data
point. Note that we consider only the widths obtained from single Gaussian fits
to the data, i.e. assuming no anisotropy of the velocity distribution. The line
profiles obtained at 3.5~$R_{\sun}$ for the different density models present a
slight flattening at the central frequency which makes their
one-Gaussian-fitting less accurate. The corresponding line widths in
Fig.~\ref{Line_Width_MgX_New271103_AlpsVar.eps} are therefore more uncertain.

At heights above $2~R_{\sun}$ the variation of the width of the synthetic
profiles of \ion{Mg}{x} as a function of the projected heliocentric distance
shows how sensitive the collisional line profile is to the details of the
electron density stratification. This explains why the profiles emitted by heavy
ions in the corona, namely \ion{O}{vi} and \ion{Mg}{x}, which are partially or
totally collisional, are larger compared to the ones of Ly-$\alpha$ (pure
radiative lines).

A comparison with Fig.~\ref{Line_Width_Ratio_New_DOYGUH_Paper.eps} shows that
the line widths obtained for the \ion{Mg}{x} lines are only slightly larger than
for the \ion{O}{vi} lines. This suggests that already for \ion{O}{vi} the
excitation through electron collisions dominates over resonant scattering.

\section{Discussion} 

We have considered the effect of the density stratification on the strongest
lines observed by UVCS in the polar holes of the solar corona in the presence of
a consistent model of the large scale magnetic field and solar wind structure.
We consider lines with different formation processes (pure scattering lines,
purely collisionally excited lines, and lines for which both processes are
important; represented by Ly-$\alpha$, the \ion{Mg}{x} and \ion{O}{vi} doublets,
respectively). We calculate the line profiles, total intensities and (for the
\ion{O}{vi} lines) intensity ratio of the spectral lines emitted at different
altitudes in the polar coronal holes. Integration along the LOS is taken into
account. The velocity distributions of the reemitting atoms/ions are considered
to be simple Maxwellians with a drift macroscopic velocity vector that is
identical to the solar wind outflow velocity, which is calculated according to
mass-flux conservation. We note that the main assumption underlying this
investigation is that no anisotropy in the kinetic temperature of the scattering
atoms/ions is considered.

It is found that whereas the width of the Ly-$\alpha$ profile reacts little to
the choice of density stratification, the widths of the \ion{O}{vi} and
\ion{Mg}{x} lines are strongly dependent. The Ly-$\alpha$ total intensities
obtained from the different density stratifications are comparable and
reasonably fit the observed intensities by most models at most heights.

The line widths of the \ion{O}{vi} and \ion{Mg}{x} doublets obtained at
different heights in the polar coronal holes are found to be very sensitive to
details of the electron density stratification. At low altitudes in the polar
coronal holes, the calculated profiles of the lines of a given doublet all have
comparable widths, which is due to the fact that the main contribution comes
from the area around the polar axis, due to the fast drop of the electron
density with height. At greater heights, the density drops more slowly and
contributions to the LOS integrated profile from sections of the LOS with large
$|Z|$ are increasingly important. At these relatively large heights, where the
solar wind speed reaches values that leave the \ion{O}{vi} line out of
resonance, the reemitted lines behave like pure collisional lines. This is
confirmed by the fact that the calculated \ion{O}{vi} and \ion{Mg}{x} lines
exhibit similar, large line widths at these altitudes. Their widths are very
sensitive to the LOS solar wind speed, which is strongly dependent on the
density stratification, so that the differences between the profiles obtained
through different density stratifications start to show up clearly at greater
heights.

The intensity ratios of the \ion{O}{vi} doublet drops from values above 2 at
heights of $1.5~R_{\sun}$ to reach a minimal value that is very close to unity
at a height that depends on the density model. For most of the considered
density stratification models, the obtained intensity ratios are within the
error bars outlining the scatter of the observations. The LOS integrated total
intensities of the \ion{O}{vi} lines also show only a weak dependence on the
density stratification. Generally, the obtained values are reasonable compared
to the observations for both lines of the \ion{O}{vi} doublet. Note that for the
total intensity the absolute values of the electron densities are as important
as their stratification, which may explain the small dependence of the obtained
intensities on the density stratification.

For the DKL density stratification the calculated parameters are close to those
obtained from observations carried out by UVCS, in spite of the fact that we
only consider an isotropic kinetic temperature of the scattering atoms and ions.
Our analysis does not aim to decide between density models, since the exact
results depend on, e.g, the details of the magnetic structure, or on how well
the contribution of the polar plumes or foreground coronal material has been
separated from the part of the coronal hole giving rise to the fast wind.
However, it shows that a reasonable combination of the magnetic, solar wind and
density structures exist that reproduce a wide variety of the observed
parameters. Our analysis confirms and strengthens the conclusion of RS04 that
the need for anisotropic velocity distributions (i.e. anisotropy of kinetic
temperature of the heavy ions) in the solar corona may not be so pressing as
previously concluded, although we stress that the current results do not rule
out such anisotropies.

It is beyond the scope of the present paper to determine the implications
of this result for the mechanisms of heating and acceleration of different
species in the polar coronal holes. In particular, it remains to be shown to
what extend the ion cyclotron waves proposed to heat the outer corona in
numerous studies (Cranmer et al. 1999c; Isenberg et al. 2000 \& 2001; Isenberg
2001 \& 2004; Markovskii \& Hollweg 2004; etc.) are consistent with an isotropic
non-thermal broadening velocity (which is known to be present in all layers of
the solar atmosphere).

We note that the difference between the kinetic temperatures of heavy ions
and protons, found earlier by Li et al. (1997), is also present in our analysis.
If, however, we consider the more realistic case that the $\alpha_s$ values
obtain a contribution both from thermal and non-thermal broadening, then the
protons and heavy ions give more consistent results. If we further assume both
protons and heavy ions to be affected by a temperature of $0.9~10^6$ K, which is
typical of the electron temperature in the polar coronal holes (David et al.
1998), we find that the non-thermal broadening for the \ion{O}{vi} lines varies
between $\sim85$ km~s$^{-1}$ at 1.5~$R_{\sun}$ to $\sim210$ km~s$^{-1}$ at
3.5~$R_{\sun}$. For Ly-$\alpha$ these values are $\sim165$ km~s$^{-1}$ and
$\sim185$ km~s$^{-1}$, respectively; i.e. they lie between the values deduced
from the \ion{O}{vi} lines. This relative consistency between the non-thermal
broadening of protons and heavy ions is all the more surprising since we did not
in any way optimize the computations with such an aim. We note that an extended
model including the influence of polar plumes (Raouafi \& Solanki, in
preparation) suggests that the difference in the non-thermal broadening at low
altitudes is at least partly due to the neglect of polar plumes in the present
computations. Once more, we stress that the current work in no way rules out
alternative interpretations; it simply opens up the possibility of considering
models in which protons, heavy ions and electrons all have the same temperature.
In agreement with earlier results we find a somewhat lower outflow speed for
protons than heavy ions, however.

\begin{acknowledgements} 
The authors would like to thank Dr. T. Holzer for the very constructive and
encouraging report on the manuscript. We are grateful to Steven Cranmer, Bernhard
Fleck, Bernd Inhester, Eckart Marsch, Klaus Wilhelm, Giannina Poletto and Jean-Claude
Vial for helpful discussions and critical comments that greatly improved the paper.
\end{acknowledgements}

\end{document}